\pdfoutput=1

\documentclass[aip,rmp,amsmath,amssymb,onecolumn,12pt,tightenlines]{revtex4-1}

\usepackage{graphicx}
\usepackage{dcolumn}
\usepackage{bm}

\renewcommand{\thesection}{\arabic{section}}
\renewcommand{\thesubsection}{\arabic{section}.\arabic{subsection}}
\renewcommand{\thesubsubsection}{\arabic{section}.\arabic{subsection}.\arabic{subsubsection}}

\makeatletter
\renewcommand{\p@subsection}{}
\renewcommand{\p@subsubsection}{}
\makeatother

\begin{document}

\title{Commissioning of the Francium Trapping Facility at TRIUMF}

\author{M.~Tandecki}
\affiliation{TRIUMF, Vancouver, BC V6T 2A3, Canada}

\author{J.~Zhang}
\affiliation{JQI, Department of Physics and NIST, University of Maryland,
  College Park, MD 20742, USA}

\author{R.~Collister}
\affiliation{Dept. of Physics and Astronomy, University of Manitoba,
  Winnipeg, MB R3T 2N2, Canada}

\author{S.~Aubin}
\affiliation{Department of Physics, College of William and Mary,
  Williamsburg VA 2319, USA}

\author{J.~A.~Behr}
\affiliation{TRIUMF, Vancouver, BC V6T 2A3, Canada}

\author{E.~Gomez}
\affiliation{Instituto de F{\'i}sica, Universidad Aut{\'o}noma de San Luis Potos{\'i},
 San Luis Potosi 78290, Mexico}

\author{G.~Gwinner}
\affiliation{Dept. of Physics and Astronomy, University of Manitoba,
  Winnipeg, MB R3T 2N2, Canada}

\author{L.~A.~Orozco}
\email{lorozco@umd.edu}
\affiliation{JQI, Department of Physics and NIST, University of Maryland,
  College Park, MD 20742, USA}

\author{M.~R.~Pearson}
\affiliation{TRIUMF, Vancouver, BC V6T 2A3, Canada}


\setcitestyle{numbers,square}

\begin{abstract}
We report on the successful commissioning of the Francium Trapping Facility at \mbox{TRIUMF}. Large laser-cooled samples of francium are produced from a francium ion beam delivered by the ISAC radioactive ion beam facility. The ion beam is neutralized on an yttrium foil, which is subsequently heated to transfer the atoms into the magneto-optical trapping region. We have successfully trapped $^{207}$Fr, $^{209}$Fr and $^{221}$Fr, with a maximum of $2.5 \times 10^5$ $^{209}$Fr atoms. The neutral cold atoms will be used in studies of the weak interaction through measurements of atomic parity non-conservation.
\end{abstract}


\maketitle

\tableofcontents

\section{Introduction}

Parity non-conservation (PNC) is a unique signature of the weak
interaction. Thanks to this property it is possible to study the weak interaction in the presence of other, dominating interactions like the electromagnetic one. Shortly after the discovery of PNC, the possibility of observing it in atomic systems, due to the possible existence of a neutral weak current, was considered \cite{zeldovich59}.

We have initiated a program, at the Francium Trapping Facility (FTF) at ISAC/TRIUMF, to study weak interaction physics at low energy through atomic parity non-conservation (APNC) \cite{bouchiat97} in neutral francium (Fr) atoms. The scarcity of the unstable atoms can be compensated for by the techniques of laser cooling and trapping applied to radioactive atoms \cite{sprouse97} that allow for long-term interrogation of the atoms under well-controlled conditions.

The APNC measurements rely on two aspects; first, an apparatus with a well-defined handedness, and second, the interference between a weak-interaction-induced transition and an electromagnetically allowed transition. The control of the environment and the apparatus require careful planning at all stages, in particular since the measurement takes place in an accelerator hall, where changing magnetic fields and large RF electromagnetic fields can introduce noise and systematic effects. The program requires a reliable laser cooling and trapping apparatus that efficiently uses the radioactive ions. While the accelerator delivers fast ions to our facility, the planned experiments require slow neutral atoms.

Our APNC apparatus consists of two parts, the capture assembly and the science chamber. The first step in the manipulation of francium is to accumulate the francium ions on an yttrium foil. The francium is released in atomic form (because of the lower work function of yttrium with respect to francium \cite{dinneen96}) by resistive heating of the foil, and it is trapped in a magneto-optical trap (MOT). All of this happens in the capture assembly.
Briefly, a MOT uses a combination of near-resonant laser light (red-detuned from resonance by $\Delta$) with a three-dimensional magnetic field gradient to produce a force that is both velocity-dependent (dissipative) and position-dependent (restorative) to cool and capture neutral atoms that are moving at velocities lower than a capture velocity, $v_c$ (about $15$\,m/s) \cite{sprouse97}.
Once trapped, the atoms can be probed in the capture assembly or transported to the science chamber for the actual APNC measurements in a well-controlled environment.

This paper reports the successful commissioning of the FTF at the ISAC facility of TRIUMF; we focus on the capture assembly, as the science chamber was not constructed yet at the time of the commissioning run. Section~\ref{sec-weak} briefly reviews the physics of APNC, followed by the requirements for the measurements and facility in section~\ref{sec:requirements}. The production and delivery of francium into the FTF is presented in section~\ref{sec:production_delivery}, while an overview of the FTF is given in section~\ref{sec:FTF}. Brief results from measurements with rubidium are presented in section~\ref{sec:rubidium}. Section~\ref{sec:commissioning} summarizes the results from the on-line commissioning run. 
Section~\ref{sec:conclusions_outlook} gives conclusions for the described work and an outlook.

\section{Weak interaction in atoms}\label{sec-weak}

Atomic parity non-conservation arises from the parity-violating  exchange  of  $Z^0$  bosons  between the atomic constituents,  leading  to  a  mixing  of  atomic  levels  of  opposite  parity  \cite{bouchiat97}.
As a result, otherwise parity-forbidden electric-dipole  transitions  can  be  excited  between  states  of  the  same  parity. Several processes lead to APNC. If the nucleons and electrons are described as as a vector and an axial vector current respectively this leads to nuclear spin-independent APNC ({\it nsi}-APNC). 
\cite{bouchiat97}.  Nuclear spin-dependent APNC ({\it nsd}-APNC) occurs in three ways \cite{zeldovich59,flambaum84};
(i) an electron interacting weakly with a single valence nucleon (nucleon axial-vector current and electron vector current),
(ii) an electron experiencing an electromagnetic interaction with a nuclear chiral current created by parity-violating weak interactions between
nucleons (anapole moment) \cite{flambaum84,flambaum97}, and
(iii) the combined action of the hyperfine interaction and the nuclear spin-independent $Z^0$ exchange interaction from nucleon vector currents \cite{haxton01,johnson03,ginges04}.
In heavy atoms, process ii) is the dominant source of {\it nsd}-APNC \cite{flambaum97,haxton01,gomez06}. 

The FrPNC collaboration is working towards measurements of both types of PNC effects \cite{gwinner06b}. The attractiveness of Fr for {\it nsi}-APNC experiments has been discussed since the early 1990s in the context of searches for new physics beyond the Standard Model (SM) \cite{gomez06,marciano90,behr09,davoudiasl12,ramsey-musolf06}.
The atomic theory and structure of Fr (Z = 87) can be understood at a level similar to that of Cs (Z = 55), where the most precise measurement to date has been performed \cite{wood97,wood99}, yet the APNC effect has been calculated to be 18 times larger \cite{dzuba95,safronova00}.

All recent and on-going experiments in atomic PNC rely on the
large enhancement of the effect in heavy nuclei (large $Z$), first pointed out  by the
Bouchiats \cite{bouchiat74,bouchiat74b,bouchiat75}. The weak interaction transition amplitudes
are exceedingly small, and an interference method is commonly used to
measure them. A typical experiment measures a quantity that has the form
\begin{equation}
|A_{\rm{PC}} \pm A_{\rm{PNC}}|^2 \approx |A_{\rm{PC}}|^2  \pm 2Re(A_{\rm{PC}}A_{\rm{PNC}}^*),
\label{interference}
\end{equation}
where $A_{\rm{PC}}$ and $A_{\rm{PNC}}$ represent the parity-conserving (typically much larger) and parity non-conserving amplitudes respectively. The second term on the right
side corresponds to the interference term, which can experimentally be isolated because it changes sign under a parity transformation.

Our strategy for APNC in Fr is to measure the excitation rate of a highly forbidden transition inside a handed apparatus (see the review by the
Bouchiats \cite{bouchiat01}). For the case of { \it nsi}-APNC it will be the electric dipole transition between the $7s$ and $8s$ levels in Fr that becomes allowed through the weak
interaction. Interference between this transition and the one induced by the Stark effect due to the presence of a static electric field generates a signal proportional to the weak charge. We follow a strategy similar to the Cs APNC measurement that reached a precision of 0.35\% \cite{wood97,wood99} on the {\it nsi}-APNC.
For the case of {\it nsd}-APNC it will be the electric dipole transition between hyperfine levels in the $7s$ ground state interfering with the allowed magnetic dipole transition to produce a signal proportional to the anapole moment. In Refs.~\cite{wood97,wood99} they also obtained the first definite measurement of an anapole moment with an accuracy of 14\% by comparing signals from different hyperfine levels.
The successful extraction of weak interaction  physics  from  the  measured  atomic  quantities  requires  a  detailed  quantitative understanding  of  the  atomic  wavefunctions \cite{porsev09,porsev10,dzuba12c}.

An APNC measurement will generally proceed as follows; we trap and cool atoms on-line at the FTF.
The cold francium sample is then transferred to a science chamber with precise control of all electric, magnetic, and electromagnetic fields, as well as better vacuum ($< 10^{-10}$\,mbar), where the handed experiments to study APNC take place. We require probing of large numbers of atoms in order to obtain a good signal-to-noise ratio. Our preliminary calculations show that $10^{5}$ to $10^{6}$ trapped atoms at a temperature below $1$\,mK should be sufficient.

\section{Requirements of the Francium Trapping Facility}\label{sec:requirements}

\begin{figure}
\begin{center}
  \includegraphics[width=0.85\linewidth]{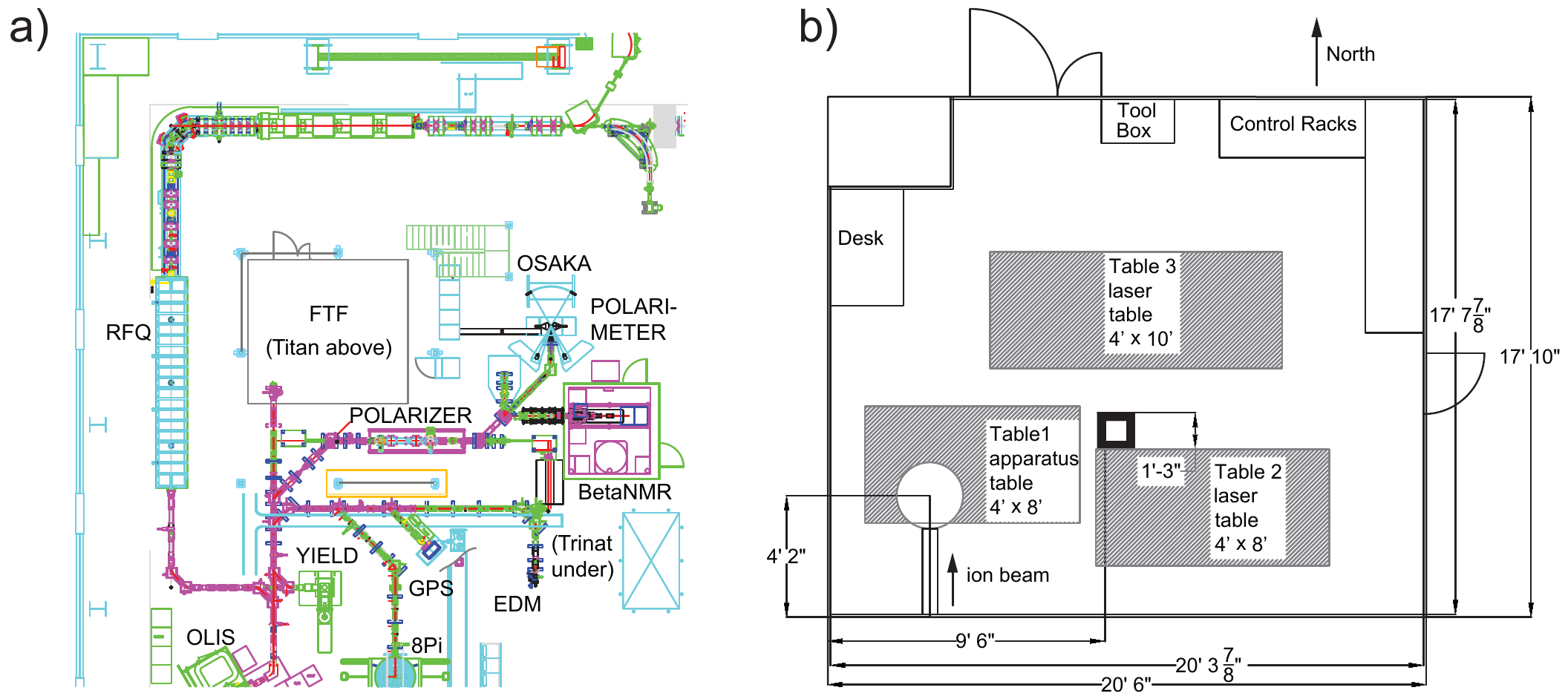}
\caption{a) FTF in the ISAC I Hall at TRIUMF. b) Layout of the FTF with the francium beamline, optics tables and control equipment in August 2012. }
\label{fig:plan}
\end{center}
\end{figure}

Reference~\cite{gomez07} has a detailed study of all the necessary requirements for the measurement of $\it nsd$-APNC in Fr. The environmental requirements are similar for the $\it nsi$-APNC measurement. The ISAC-I Hall at TRIUMF, where our experiment resides, has several linear accelerators, or linacs (see for example the beamline denoted RFQ in figure~\ref{fig:plan}a),  that produce ample radio-frequency (RF) electromagnetic fields that can interfere with our measurements.
The complete FTF is therefore located inside a Faraday cage.

\subsection{RF shielding and temperature environment}
\label{sec:ftf_room}

The background RF noise at the location of the FTF without the Faraday cage was 21 V/m at 35 MHz and 2.7 V/m at 106 MHz, the most relevant frequencies from the RFQ of the linac on the west side of the FTF (see figure~\ref{fig:plan}a).  The dimensions of the Faraday cage are $5.9 \times 6.25 \times 3.66$\,m$^3$ under the existing platform that houses the TITAN facility. There is a column that supports the TITAN facility so the shielding of the room has to cover the column appropriately. Universal Shielding Corporation designed and constructed a galvanized steel Faraday cage to be the laboratory of the FTF. The floor, ceiling, and walls are made of metal-covered wood composite, with the appropriate fire retardants. The doors required double sets of grounding strips  to ensure no RF leakage into the room. The room has penetrations with RF filters to allow water, air, the beamline, telephone, Ethernet, and other signals and services to enter the room. When the room is properly closed, the Faraday cage attenuates the RF amplitude by more than 100 dB at $35$\,MHz to levels lower than the noise floor of our measuring instrument.

The requirements on the room temperature are a range of 22-24 ${^\circ}$C and a stability better than $\pm$1.0 ${^\circ}$C or better (year round, minute-to-minute). The relative humidity must be stable to $\pm 5$\,\%. The temperature and relative humidity of the ISAC area of TRIUMF show short-term fluctuations of $\pm  3~{^\circ}$C and $\pm$ 15\%, respectively, while the long-term ones are season-dependent.
Most of the equipment in the FTF requires this level of environmental stability, in particular the lasers and their locking systems \cite{zhao98}.
A few of the systems have extra temperature control as their stability should be better than $0.1^{\circ}$C. The temperature and humidity stabilization is realized by an HVAC system. 


\subsection{Magnetic field environment}

The applied magnetic fields at the center of the science chamber need a fractional control at the $10^{-5}$ level \cite{gomez07}, with applied fields of the order of a few gauss initially.
During the summer of 2012, with most of the trapping apparatus (capture assembly) assembled and working, we studied the magnetic field environment of the FTF and measured the spectral density of the magnetic field noise. We measured the magnetic field with a Honeywell HMC1002 Anisotropic Magneto-Resistive (AMR) sensor, which has a precision of 100 $\mu$G, with a 10 Hz bandwidth and a range of $\pm$2 G.
The magnetic noise at the experiment is dominated by line noise at 60 Hz and 180 Hz at the mG level that can be controlled with feedback and feedforward.

The magnetic field caused by the TRIUMF cyclotron at the location of the FTF is of the same order as the magnetic field of the Earth, $\sim 0.5$\,G. Above the FTF sits the TITAN facility with superconducting magnets for their Penning traps. Their influence at the MOT is also of the order of the magnetic field of the Earth, and since the magnets operate in steady state, it is possible to correct for the DC fields with appropriate pairs of coils surrounding the apparatus. 
Measurements before the fall of 2011 at the current location of the FTF indicated that the DC magnetic field varies on the order of $100$\,mG over the course of several days. While we will not shield the entire room against quasi-static magnetic fields,  active control at the science chamber will be necessary to maintain a highly-stable magnetic field environment for precision APNC  measurements.

\subsection{Radiation safety}
The overall design of the FTF incorporates a number of radiation safety features. Radiation safety is an important consideration for experiments at TRIUMF and also a part of standard engineering practices.
We briefly describe these standard nuclear physics engineering steps and procedures in the context of francium trapping, since they are not usually encountered in atomic physics experiments.

The FTF is equipped to handle the short-term radiation present during delivery and trapping of francium and also the long-term radiation hazards posed by the francium decay products inside the vacuum system. At present, the radiation levels during francium delivery and trapping are sufficiently low ($< 10$\,$\mu$Sv/hr) to permit in situ operation of the FTF by experimenters, though the computer control of the apparatus can be used for future external control of the FTF if higher francium and radiation rates become necessary. 
The FTF room is designed for connection to ISAC's nuclear exhaust system in case of an accidental vacuum breach during a beamtime or a planned opening of the vacuum system. The turbomolecular pumps on the beamline are connected to this nuclear exhaust as well.
Several francium isotopes decay to long-lived isotopes. For example, $8.7$\,\% of $^{209}$Fr, used during the commissioning run, decays into $^{209}$Po, which is an $\alpha$ emitter with a half-life of $102$\,years. Handling components that are contaminated with these isotopes requires special care as the $\alpha$ radiation is hard to detect and is very damaging to biological tissue when inhaled or ingested (compared to $\beta$ or $\gamma$ radiation). The vacuum system components, in particular the neutralizer chamber (see section~\ref{sec:neutralizer_chamber}), have been designed for rapid swap-out in order to minimize exposure to these isotopes.

During the run, spikes in the radiation levels were detected by $\gamma$ monitors close to the roughing lines of the turbomolecular pumps. The spikes ($> 10$\,$\mu$Sv/hr) coincided with the heating of the yttrium foil and indicate that significant amounts of activity are pumped out of the vacuum system into the nuclear exhaust. After opening the system, we detected, by $\gamma$ counting, a relatively large amount of $^{209}$Po in the yttrium foil (corresponding to $>50$\,\% of delivered francium) and a small amount in the glass cell (where only upper limits could be set). The reason for the activity in the yttrium foil is that not all francium atoms are released from it during heating, see section~\ref{sec:efficiency}.


\section{Francium production and delivery}~\label{sec:production_delivery}

In this section we briefly describe the production of the francium ions, and their transport to the FTF. 

\subsection{Production of francium at TRIUMF}

\begin{figure}
\begin{center}
  \includegraphics[width=0.75\linewidth]{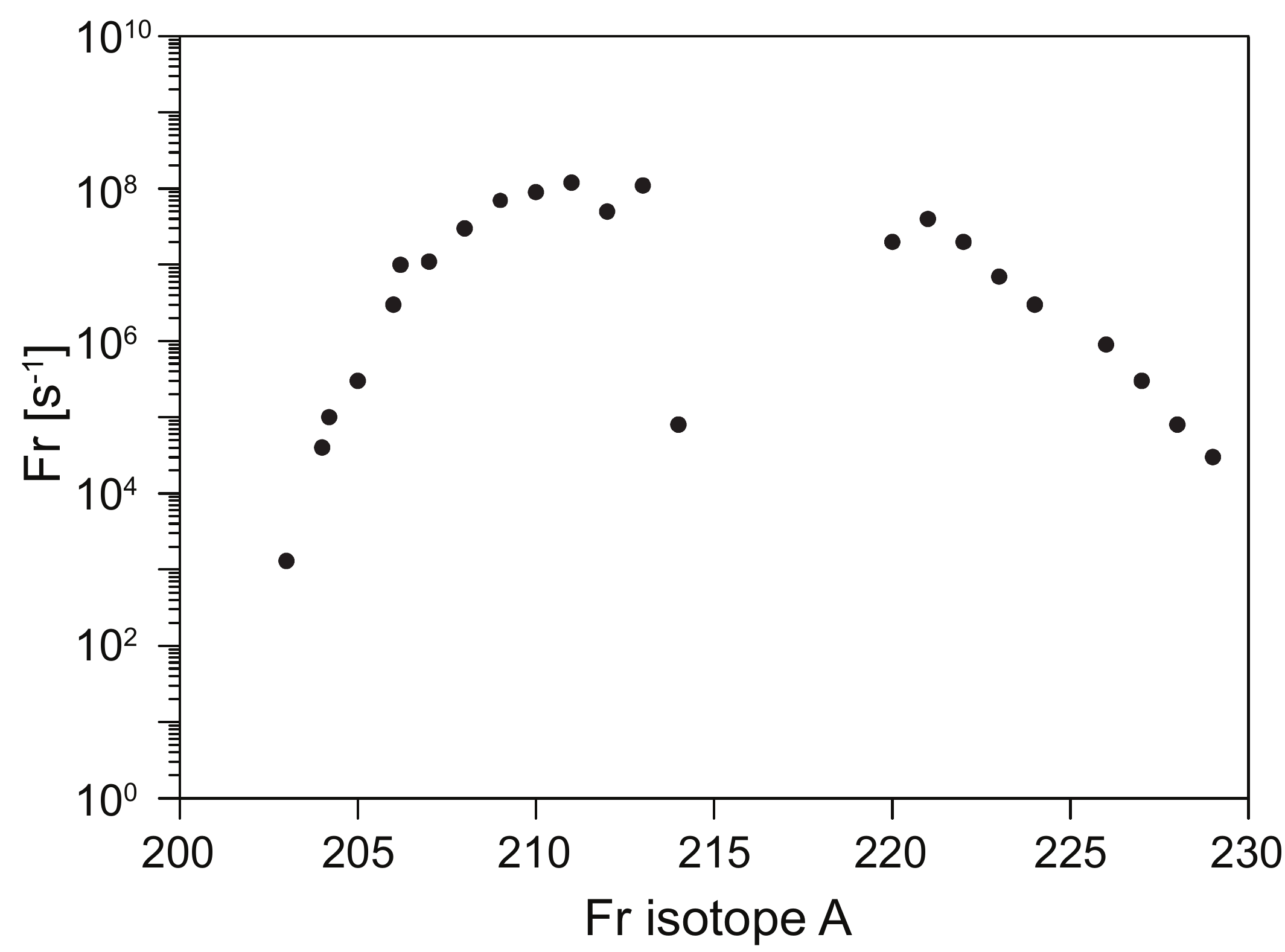}
\caption{Production yield of Fr isotopes at the ISAC facility at TRIUMF with $2$\,$\mu$A of protons on a UCx target in December 2010. Yields have increased an order of magnitude since then.
}
\label{fig:yield}
\end{center}
\end{figure}

A  proton beam from the TRIUMF cyclotron with an energy of $500$\,MeV impinges on a target and produces a number of neutron-rich and neutron-deficient Fr isotopes.
The proton current can be as high as $10$\,$\mu$A for UC$_x$ targets.
These isotopes originate from the interaction of the protons with a hot UC$_x$ target made of many thin wafers of the material; see Ref.~\cite{dombsky00} for information about ISAC. The produced Fr isotopes are surface-ionized in a hot transfer tube, pass through a mass separator that selects a single nuclear mass and are guided through ISAC's low-energy beamline. The production yields of about $1 \times10^8$ Fr/s demonstrated in December 2010 for several francium isotopes have been increased during the two beamtimes in 2012. 
Although isobaric contaminants may be delivered to the experiment, these unwanted atoms are not trapped, since the MOT employs narrow isomer-specific optical transitions. figure~\ref{fig:yield} shows the yield of Fr isotopes available during the commissioning of the UC$_x$ target in December 2010.

\subsection{Beamline of the FTF}\label{sec:beamline}

 \begin{figure}
\begin{center}
  \includegraphics[width=0.80\linewidth]{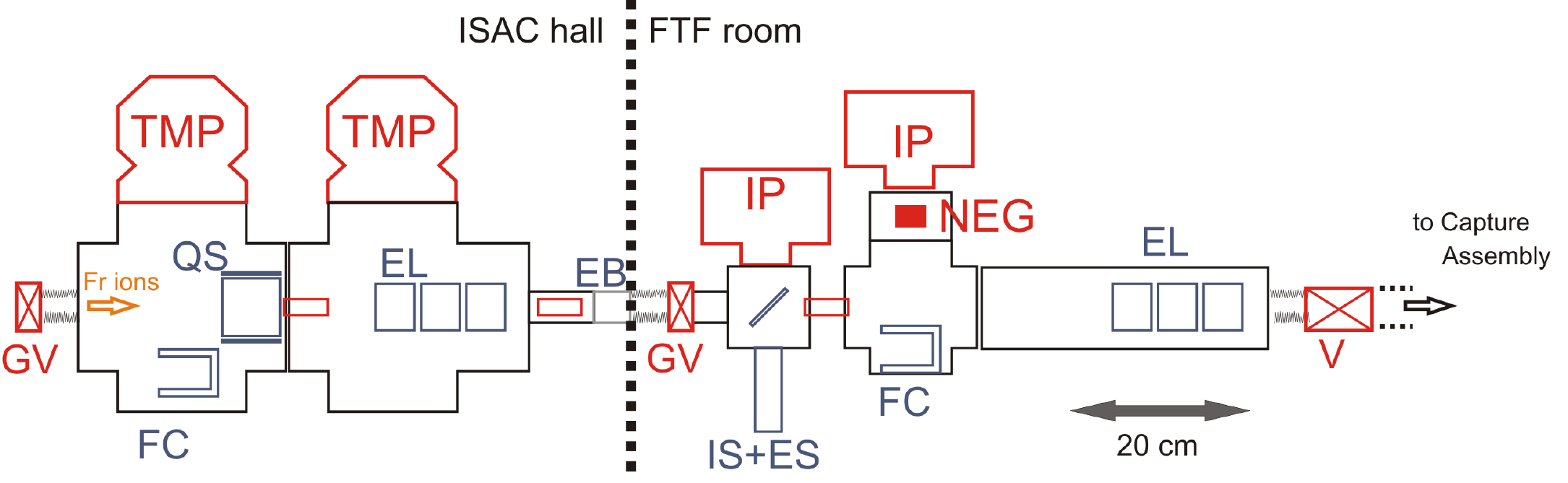}
\caption{An overview of the beamline delivering the francium ions from the ISAC-I low-energy beamlines to the capture assembly. Part of the beamline is located outside of the room, and part is inside. There is an electric break in the beamline so that the room is not grounded to the outside beamline. The vacuum system is color-coded in red, the beam delivery in blue. The abbreviations used in the figure are the following; EB: electric break, (G)V: (gate) valve, TMP: turbomolecular pump, QS: quadrupole elements + electrostatic steerer, EL: einzel lens, FC: Retractable Faraday cup, IP: ion pump, IS+ES: ion source + retractable electrostatic mirror, NEG: non-evaporative getter cartridge. The capture assembly together with the final differential pumping diaphragm (inner diameter of $9.5$\,mm) are located to the right of the final valve. Bellows (staggered lines in the figure) link different pieces of the beamline. }
\label{fig:elec-vacuum}
\end{center}
\end{figure}

All the focusing and steering elements of the beamline that deliver Fr$^{1+}$ onto the yttrium neutralizer foil (see section~\ref{sec:neutralizer_chamber}) 
are electrostatic for mass-independent ion transport. We have the possibility to use the same beamline to feed in stable isotopes such as Rb$^+$ from a commercial ion source (Kimball Physics ILG-6) to align and test the full system. 
Four steerer plates and two einzel lenses control the ion beam from ISAC to the neutralizer chamber with the yttrium foil. Furthermore, the 4-sector x-y steerer plates can be independently controlled to add a quadrupole singlet to optimize the final tune after the astigmatic ISAC beam transport design \cite{baartman03}.
For a $20$\,keV beam, the voltages on the einzel lenses are of the order of $+12$\, kV, while the voltages on the steering plates are between $\pm 300$ and $\pm 400$\,V.

We ran a test  in January 2012 with a stable $^{16}$O$^{1+}$ ion beam from ISAC at $20$\,keV which is similar to the energy used for Fr$^{1+}$. 
We delivered $90$\% of the beam in a location about 20 cm before the neutralizer foil -- before the capture assembly was installed -- through a 6 mm aperture. The locations of the electrostatic elements and beam diagnostics are indicated in figure~\ref{fig:elec-vacuum}. Stable $^{238}$U$^{1+}$ beam coming out of the target is routinely used at ISAC for tuning the beamline optics prior to Fr delivery.

The vacuum inside ISAC low-energy beamlines is of the order of $10^{-7}$\,mbar, while we require a vacuum in the capture assembly that is three orders of magnitudes lower. The first section of the FTF beamline is pumped by two $550$\,l/s turbomolecular pumps (Varian Turbo-V 551 Navigator), which are located outside of the Faraday cage. Combined with differential pumping diaphragms (with a diameter of $25.4$\,mm and a length of $50.8$\,mm), they bring down the pressure by two orders of magnitude. The second section, inside the Faraday cage, is pumped by two $30$\,l/s ion pumps (Duniway DGD-050-5143-M) and a non-evaporable getter (NEG) cartridge (SAES WP 38/950-ST 707, with a $430$\,l/s pumping speed for H$_2$) and is separated from the capture assembly by a tube with a diameter of $9.5$\,mm and a length of $19$\,mm. The locations of the different components of the vacuum system is shown in figure~\ref{fig:elec-vacuum}.

\section{Francium Trapping Facility} \label{sec:FTF}

The FTF consists of the Faraday cage (see section~\ref{sec:ftf_room}) that contains all the experimental equipment required to carry out the APNC measurements in the future. 
Some of the most important components of the facility will be described in this section; the capture assembly for trapping the francium ions into a MOT, the associated lasers, and the control and acquisition software.

\subsection{The capture assembly}

\begin{figure}
\begin{center}
  \includegraphics[width=0.85\linewidth]{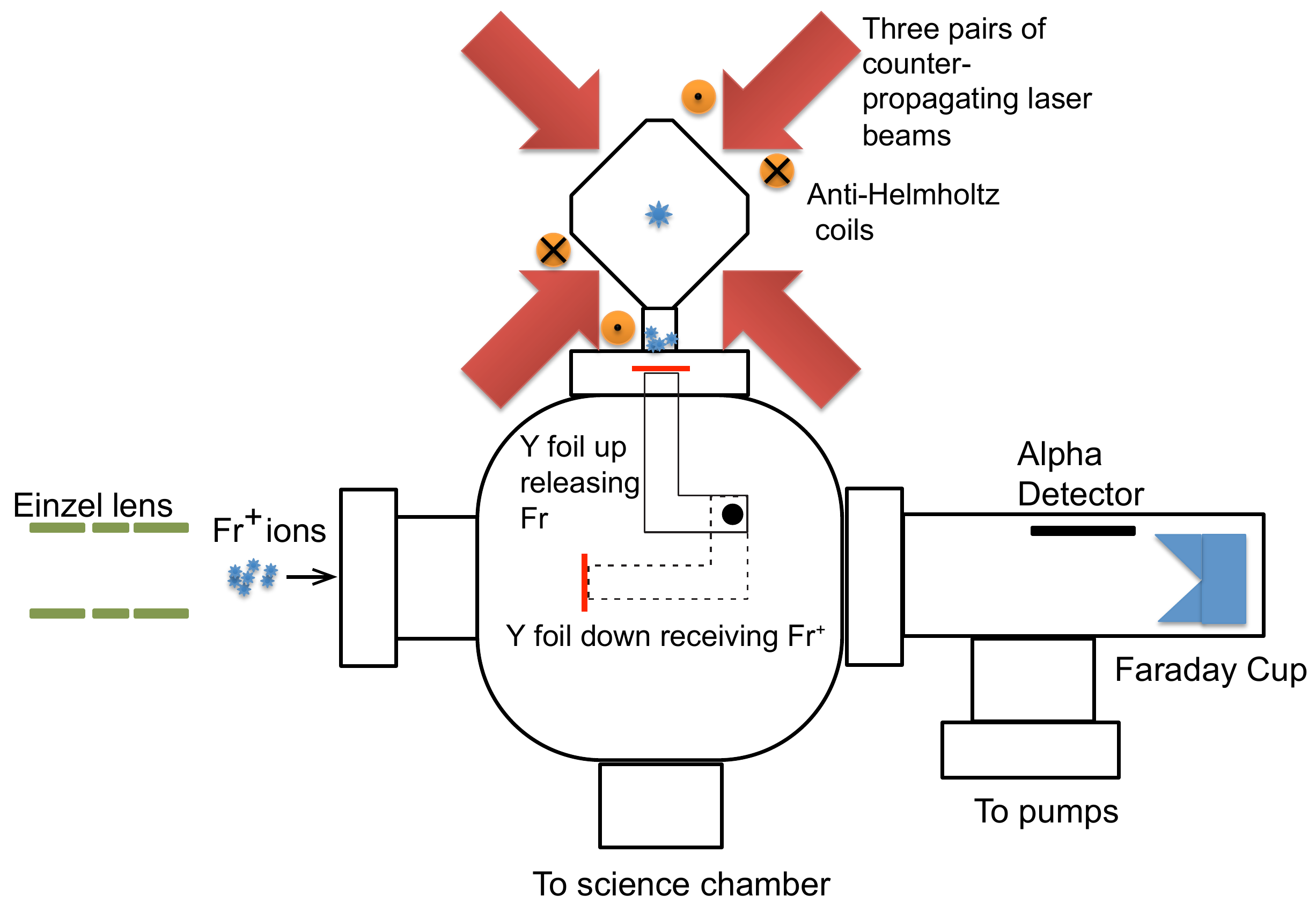}
\caption{Schematic of the capture assembly. The Fr$^{+1}$ ions impinge on the yttrium neutralizer. After accumulation, the foil is rotated and heated to release the francium atoms into the glass cell where they are trapped in the MOT. When the foil is in the upper position, the francium ions impinge on the beam diagnostics system at the end of the beamline.
}
\label{fig:MOT_sketch}
\end{center}
\end{figure}

The capture assembly is connected to the beamline and sits about $1.6$\,m above the floor. It consists of three main parts: the neutralizer, the glass cell for the MOT, and the diagnostics with a Faraday cup and $\alpha$ detector (figure~\ref{fig:MOT_sketch}). A stainless steel chamber (Kimball Physics) houses the collimator and the dual-position neutralizer.
The collimating aperture of $9.5$\,mm diameter (not shown in figure~\ref{fig:MOT_sketch}) is held on a groove grabber (Kimball Physics) at the entrance of the chamber and is used for tuning the ISAC ion beam.

The high-efficiency neutralization and capture scheme developed at Stony Brook \cite{aubin03a} has been used as a starting point for the design of the Y neutralizer. There are two important improvements on the previous design. All related parts are mounted on one 4.5-inch conflat (CF) flange and attached on one side, as to mitigate possible safety concerns related to the decay products of francium if we have to replace the neutralizer.
The current to the yttrium foil is now delivered by fixed contacts instead of the continuous wire used before that was the dominant failure mode (when breaking) in the previous design. The design allows for the atoms to be transferred from the capture MOT to the science chamber below.

\subsubsection{Neutralizer chamber}
\label{sec:neutralizer_chamber}

\begin{figure}
\begin{center}
\includegraphics[width=0.85\linewidth]{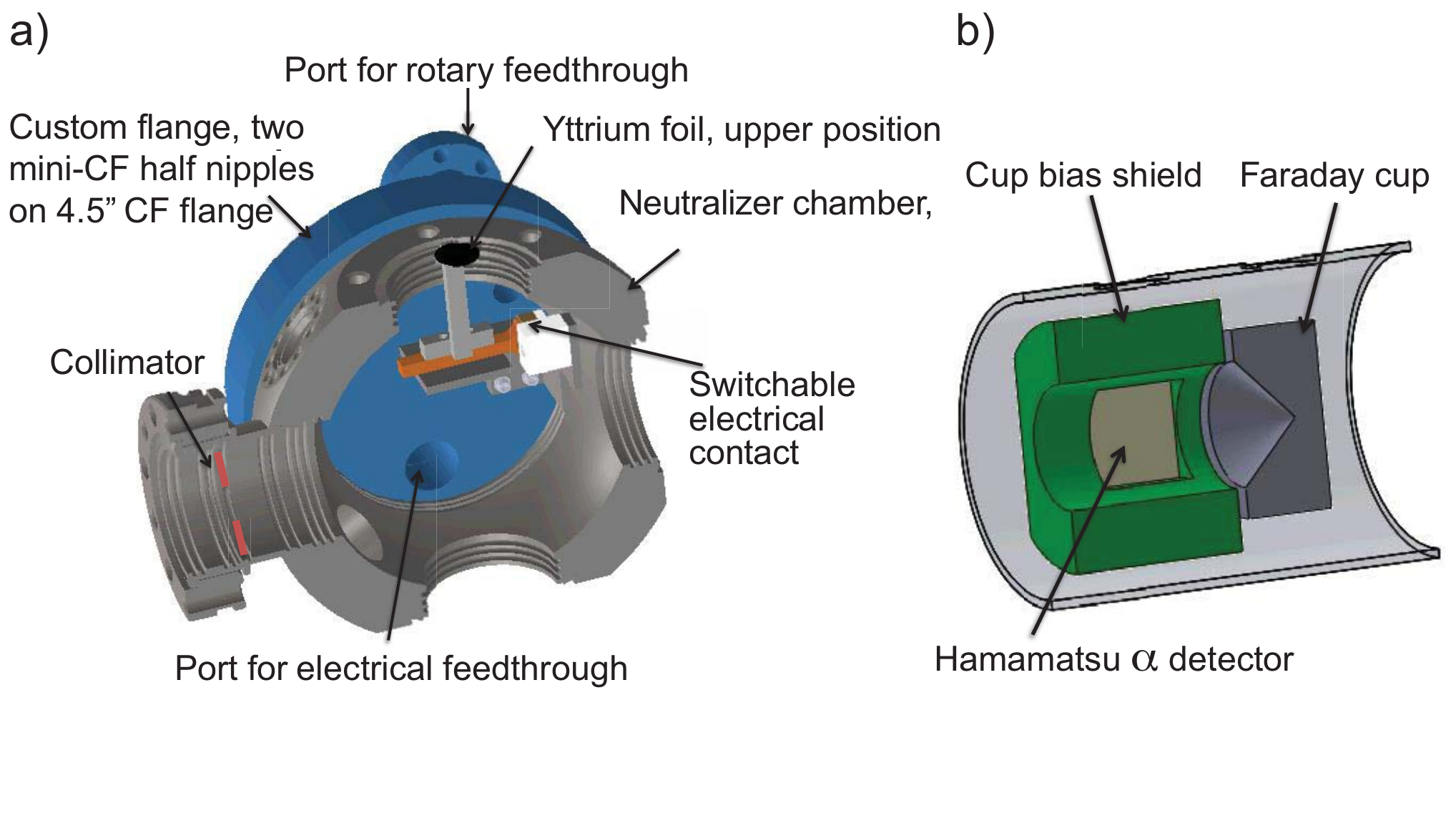}
\caption{Fr+ ion beam accumulation and detection in the capture assembly. a) Cut-away view of the neutralizer chamber.b) Cut-away view of the Faraday cup and $\alpha$ detector assembly.}
\label{fig:alpha_det}
\end{center}
\end{figure}

The ion beam is focused on the Y foil (with a thickness of $25$\,$\mu$m) in the catching position (see figure~\ref{fig:MOT_sketch} and figure~\ref{fig:alpha_det}a).
After accumulating the radioactive ions (for about $20$\,s) the neutralizer holder rotates to its delivery position and a current runs through the thin foil raising its temperature to not more than $700^{\circ}$C by running a current of about $8.9$\,A. 
The chamber has a window mounted on one of the mini-CF ports at the top of the chamber that allows observation of the neutralizer holder to measure the temperature using a pyrometer and ensure its proper rotation. Two other mini-CF ports at the bottom have electrical feedthroughs installed for the collimator read-out and for the Rb dispensers (SAES) that provide neutral Rb atoms to align and test the MOT.  
The neutralizer up position seals the bottom of the MOT glass cell.
The figure does not show a rotatable Ta foil, also on a pneumatic feedthrough, that we have used to accumulate $^{225}$Ac that decays into $^{221}$Fr \cite{tandecki13}. The Ta foil and its mount can be removed simply by taking the flange out. A combination of two pumps maintain the vacuum in the chamber: a 20 l/s ion pump (Duniway) and an ion-getter pump (NexTorr by SAES), that has 100 l/s pumping speed for hydrogen.

We are in the process of adding hardware interlocks that will monitor the position of the neutralizer holder and the Ta foil and the intensity and duration of the current pulse applied to the Y foil so as to prevent damage to the foil. The interlocks 
will not require software control and they will default to a safe state in case of a power or hardware failure.

\subsubsection{The MOT for capturing Fr}

The atom trap itself is situated in a glass cell with a silane-based dry-film coating to avoid sticking of the Fr atoms to the walls, while providing many atom-wall interactions that re-thermalize the Maxwell-Boltzmann distribution of the atoms at each `bounce'.  These numerous 
`bounces' provide multiple opportunities to capture the atoms in the MOT \cite{monroe90}. We follow a procedure based on Ref.~\cite{fedchak97} using the silane-based compound SC-77, with modifications established by trial-and-error over the years. 

The pulsed-mode operation of the trap has many benefits; by keeping the trapping cell open most of the time, we ensure a good vacuum resulting in a long trap lifetime. Closing the trap with the neutralizer delivers the neutral atoms into the trapping region and prevents them from escaping before they are trapped. Having the yttrium cold most of the time reduces damage to the dry-film coating. Finally, the francium is accumulated on the yttrium foil rather than in the MOT, thus reducing losses.
After collecting a laser-cooled sample, the atoms will then be transferred vertically down to a second chamber for the planned precision experiments. Preliminary tests using stable Rb with an off-line trap show excellent transfer efficiency (50 \%), see Ref.~\cite{sheng10,sheng12, perezPhD}; the atom cloud is pushed downwards by a short ($\sim2$\,ms), on-resonance laser pulse of $0.5$\,mW, while the trap light for the MOT is turned off and the repumping light remains on. A `counter push' is not required to successfully trap the atoms in the second chamber.

A pair of water-cooled coils provide an axial magnetic field gradient of 7 G/cm for regular MOT operation, but are capable of up to 20 G/cm. We use additional Helmholtz coils to compensate for environmental magnetic fields. The ion pump produces a significant field (about $3$\,G and a gradient of $0.3$\,G/cm) at the trap despite our use of mu-metal shielding around the magnet of the ion pump. 

\begin{figure}
\begin{center}
  \includegraphics[width=0.85\linewidth]{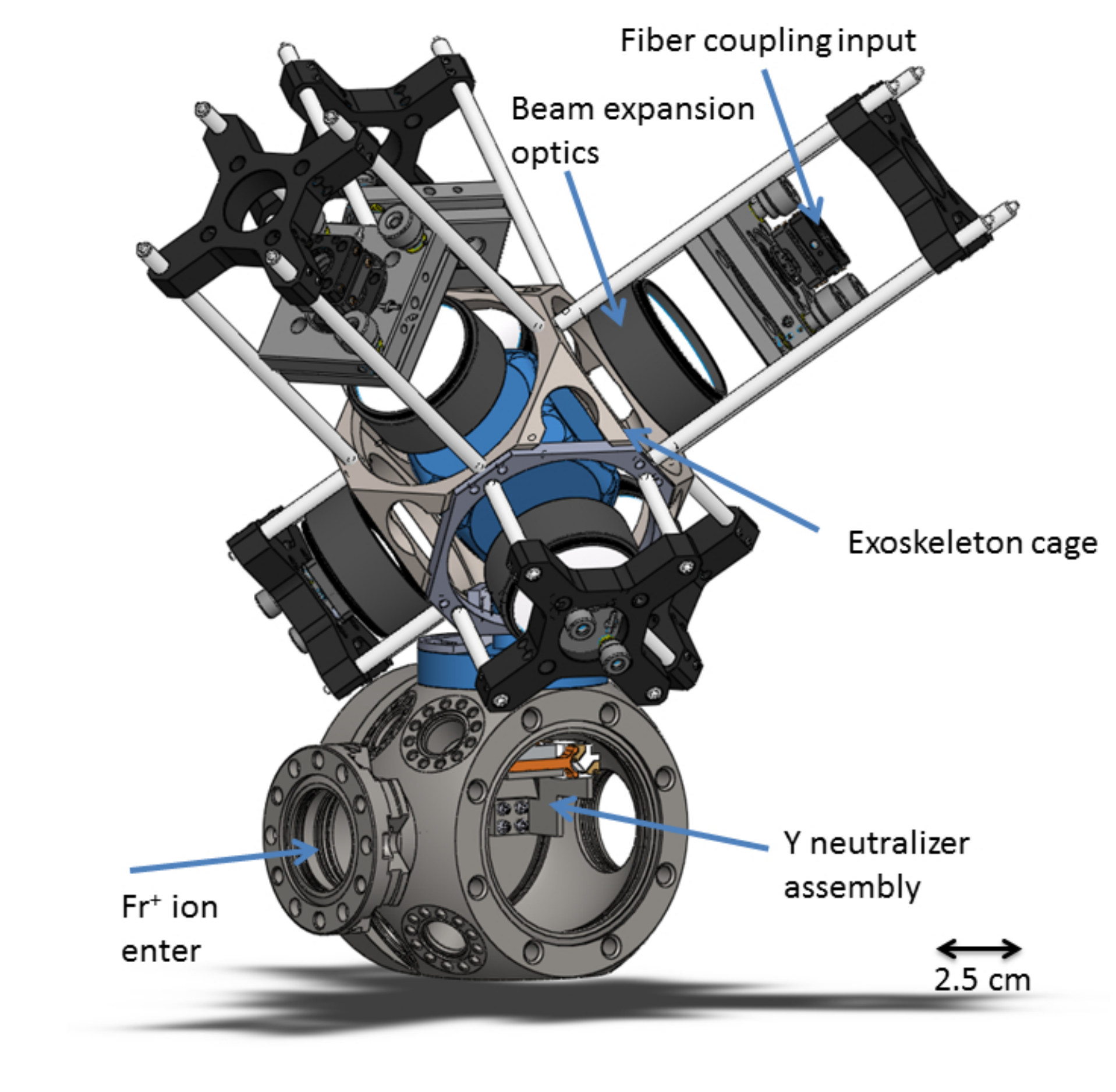}
\caption{Drawing of neutralizer chamber with the glass cell (highlighted in blue) for the MOT and the required optics. Not shown are the magnetic field coils. The size of the larger CF flanges is 4.5 in. }
\label{fig:mech_capture}
\end{center}
\end{figure}

Figure~\ref{fig:mech_capture} shows  the schematic of the capture trap with the MOT glass cell mounted on top of the neutralizer chamber, surrounded by the optics that deliver the laser beams on each side of the glass cell cube. Three identical sets of optics expand optical-fiber-delivered trapping light into $5$\,cm diameter beams; large beams are required to maximize the trapping efficiency \cite{aubin03a}. The beams are collimated and circularly polarized. A further three sets of optics retro-reflect the laser light as required for a MOT.  All the optics are mounted using commercial opto-mechanics (Thorlabs cage system). The opto-mechanical systems are supported by a stainless steel structure (exoskeleton) that ensures alignment of all the optics while protecting the glass cell. The polarization elements at the trap are optimized for 718 nm, resulting in a minor mismatch when operating with Rb.
The data acquisition (i.e. cameras and photomultiplier tube (PMT)) and laser systems are described in section~\ref{sec:camera_control} and section~\ref{sec:lasers} respectively.

\subsubsection{Faraday cup and $\alpha$ detector}
\label{sec:fcup_alpha_det}

Crucial for diagnostics before and during an on-line run is the Faraday cup at the end of the beamline shown in figure~\ref{fig:MOT_sketch} and figure~\ref{fig:alpha_det}b. The Faraday cup allows tuning the ion optics with stable beam. The tuning is done remotely in the ISAC control room; electrostatic elements and beamline diagnostics can be controlled and read back through the TRIUMF EPICS control system \cite{keitel99isacepics}. A suppression electrode before the cup can be biased to prevent unwanted escape of backscattered electrons. The francium yields are sufficiently high that the radioactive ion beam intensity can also be measured by the Faraday cup.
Care must be taken when doing that, since the radioactive decay will distort the readings of the Faraday cup. Charged particle emission ($\alpha$ and $\beta$ radiation) will induce a current after a sufficiently strong radioactive source has been accumulated. $\beta^-$ particles will result in an extra positive current, while $\alpha$ particles will give an additional negative current. 
Next to it, but not looking into the ion beam, is a silicon photodiode (Hamamatsu S3590-09) for $\alpha$ detection. The detector reading increases every time the Y foil is in the upper position, and we use it as an extra monitor for the francium beam (see figure~\ref{fig:alpha_activity}).
The solid angle of particles emitted from the center of the Faraday cup onto the $\alpha$ detector is $\sim 0.176$\,sr ($1.4$\,\%). The information of the $\alpha$ decay allows identification of elements and isotopes.

\subsection{Laser systems}
\label{sec:lasers}
The cooling, trapping, and spectroscopy lasers reside on two separate optical tables in the laboratory and are all locked to a scanning Fabry-Perot cavity \cite{zhao98}. There are two titanium-sapphire lasers (Coherent 899-21, 899-01) pumped by one $18$\,W pump laser (Coherent Verdi V18). Another titanium-sapphire laser (M~Squared SolsTiS), pumped by a $10$\,W laser (Coherent Verdi G10), was not yet fully commissioned at the time of the run. It will, however, serve as an efficient and reliable trap laser for future runs in the FTF, delivering about $1.7$\,W at $718$\,nm and $2.6$\,W at $780$\,nm. Furthermore there are two  diode lasers (Toptica DL100 and Sacher) at $780$\,nm and $817$\,nm respectively. The relevant frequencies for Rb are $780$\,nm (D2 line) and $795$\,nm (D1 line), for francium they are $718$\,nm (D2 line) and $817$\,nm (D1 line).


\subsubsection{Frequency determination and stabilization}

Francium does not have any stable isotopes, and a straightforward frequency reference to lock the lasers is not available. At this point we use a two-step approach. Firstly, a wavemeter (Bristol 621A) provides an approximate frequency determination. 
It has an accuracy of $60$\,MHz, while for trapping an absolute accuracy of $5$ to $10$\,MHz is required. Secondly, to lock the lasers during a run we use a scanning Fabry-Perot cavity, once the correct frequencies for trapping have been found \cite{zhao98}.
The cavity locking scheme is described in more detail in section~\ref{sec:cavity_lock} as part of the software controls.

\subsection{Experimental control and acquisition}
\label{sec:control_software}

At the time of the commissioning run, three main programs were available to run the experiment: a sequencer to control the experimental cycle (section~\ref{sec:cycle_control}), a program to acquire data from the various imaging systems (section~\ref{sec:camera_control}) and a program to frequency-stabilize the lasers to the transfer cavity (section~\ref{sec:cavity_lock}). The programs for experimental control and acquisition are written in LabVIEW or LabWindows. All control is done from PC computers inside the Faraday cage. They are connected to the outside through optically-isolated connections to prevent RF and ground contamination. 

\subsubsection{Sequencer}
\label{sec:cycle_control}

The timing unit of the sequencer is an FPGA  (PulseBlaster PB24-100-32k from SpinCore) that is used to send out digital triggers to devices directly or to trigger analog signals (NI-PCI 6713 cards). It was created originally by other research groups for atomic physics studies \cite{porto13}.
This system is capable of complicated sequences of pulses and voltages that will be required in the future. However, during the commissioning run, the timing sequence of the experiment was simple enough to be generated by two digital delay generators (DG535 from Stanford Research Systems).


\subsubsection{Data acquisition and camera operation}
\label{sec:camera_control}
Given the importance of images in laser trapping, we have several cameras available to acquire images of the atom cloud.
Two charge-coupled device (CCD) cameras (Roper Scientific RTE/CCD-1300-Y/HS and Imi-tech IMB-42FT) were used during the commissioning run to image the trap and to measure the number of atoms and the lifetime of the trap. The Imi-tech camera uses a commercial zoom lens (Computar MLH-10x) to give a broad field-of-view to locate the trap more easily. The Roper Scientific camera uses a custom-made double relay 1:1 imaging system with achromatic lenses.
The latter camera allows to bin several pixels together to reduce the noise and enhance the detection of small atom clouds.

A third, 1:1, imaging system has been installed since the commissioning run. It was designed with the help of commercial ray-tracing software (OSLO).
The system includes a holder to insert optical filters in the parallel section of the first relay and also a variable aperture between relay sections. 
A beam splitter at the end of the imaging system sends $30$\% of the light to a CCD camera (Micropix M-640) to ensure proper alignment of the field aperture while the remaining $70$\% reaches a photomultiplier tube for sensitive spectroscopic measurements.

We have adapted an available program for images, initially developed at the U. of Toronto \cite{aubin05}, to capture and process the optical fluorescence images from trapped atoms. 
This program can operate the Imi-tech and Micropix cameras; it can handle any FireWire-compatible camera. The Roper Scientific camera uses a program developed in our group based on LabVIEW drivers provided by the manufacturer.


\subsubsection{Transfer cavity lock}
\label{sec:cavity_lock}
The absence of readily-available atomic references at the exact frequencies necessary for the francium trapping requires the use of a transfer lock technique. We have implemented, based on previous work in Ref.~\cite{zhao98} a way to transfer the frequency stability of a HeNe laser (Melles-Griot 05-STP-901) to other lasers using a scanning, confocal Fabry-Perot interferometer. The free spectral range of the cavity is $300$\,MHz with a finesse of $100$ at $718$\,nm. Analog input and output is controlled by two National Instruments cards (PCI-6221 and PCI-6713).

The program sends a voltage ramp to a voltage amplifier (Exfo RG-91), which is then sent to one (or both) piezo crystal(s) of the cavity assembly. Light from the HeNe and the lasers requiring stabilization will be transmitted through the cavity at specific voltages of the ramp. After the cavity the light from each laser is separated either by frequency using dichroic filters or by polarizing beam splitters, and is detected by photodiodes.
The photodiode signals are acquired in synchronization with the voltage ramp 
and the relative frequency of the lasers with respect to the HeNe peaks is kept constant using a software-based PID feedback algorithm. At the time of the commissioning run everything was set up to stabilize four lasers. The feedback bandwidth is of the order of $20$\,Hz, dominated by the voltage ramp ($\sim 25$\,ms) duration, with only a small computational overhead. The stability of the lock is of the order of $\pm 5$\,Mhz over a few days. 
This is mainly limited by (i) the frequency stability of the HeNe, (ii) non-linearities in the expansion of the piezo crystals with respect to the applied voltage, and (iii) changes of atmospheric conditions, thus influencing the optical length of the cavity by changes of the refractive index of air between the mirrors. The last factor has been corrected recently by operating the transfer cavity under vacuum.

\section{Tests with stable R\lowercase{b}}
\label{sec:rubidium}

Testing the trapping apparatus with stable Rb has been crucial for understanding its performance. These tests allow us to verify the proper working of the MOT, and to align all the cameras and detectors. The exoskeleton provides good alignment of the MOT beams, but cameras (or probe beams for future measurements) are installed independently of it. 
An important test is to ensure that the zero of the magnetic field and the location of the trap coincide as much as possible. We have found that the Fr trap can appear slightly off from the Rb trap, but not more than two mm.

In the future, as the sequences for the experiment become more complicated, tests with stable Rb will become more and more important. 
They are useful to verify measurement schemes and to quantify systematic effects without using radioactive beamtime.


\subsection{Off-line Rb$^{+}$ source}
We installed a Kimball ILG-6 Rb$^{1+}$ source for delivering an off-line ion beam to test the ion optics and the neutralizer. A removable electrostatic mirror deflects the Rb$^{+}$ beam towards the neutralizer.
However, when sending an ion beam across, we found that its contribution to the atoms trapped is indistinguishable from that coming from the neutral gas background introduced by the ion source.
The distance between the ion source and the neutralizer is less than $100$\,cm, while at Stony Brook the distance to the Rb source that produced the ions was closer to $10$\,meters and included a few stages of differential pumping. There is no plan to remedy this issue, because the beam optics have been tested sufficiently in the meantime and the neutralizer scheme is working reliably.

\section{Commissioning run with F\lowercase{r}}\label{sec:commissioning}

We had five twelve-hour shifts between September 2nd and 5th, 2012 for commissioning. The isotope of choice for the test was $^{209}$Fr, as the wavelengths of the lasers necessary for trapping that isotope are well known from previous work at Stony Brook \cite{gomez06,grossman99}, it is produced at relatively high rates ($\sim 10^{9}$ Fr/s), and it minimizes long-lived $\alpha$-emitting progeny. ISAC delivered $10^6$--$5 \cdot 10^8$ Fr$^{1+}$ to the neutralizer, depending on attenuators inserted in the early stages of the beam transport. We monitor the activity by looking at the $\alpha$-decay of the atoms on the Si detector next to the Faraday cup.

Table \ref{trap_parameters} summarizes the parameters of the trap. Three of them have not been optimized yet; the laser power and stability, and the Y foil temperature. The latter was heated conservatively during commissioning.

\begin{table}[h]
\caption{Operating parameters for the September 2012 commissioning run.}
\begin{center}
\begin{tabular}{|lccc|}
\hline
Parameter    &  Value     &      Range    &   Optimized      \\ \hline
Magnetic gradient & 10 G/cm     &     $\pm$ 5 G/cm    &    yes       \\
Trapping intensity & 5 mW/cm$^2$     &    $ \pm$ 5 mW/cm$^2$    &    no  \\
Repumping intensity & 5 mW/cm$^2$     &     $\pm$ 5 mW/cm$^2$    &    no  \\
Frequency stability  &  1 MHz & $\pm$ 5 MHz/day  & partially   \\
Yttrium foil temperature & T$<700 ^{\circ}$C &  $\pm$ 50  $^{\circ}$C  & no   \\
Vacuum & $< 10^{-9}$ mbar &     & \\ \hline
\end{tabular}
\end{center}
\label{trap_parameters}
\end{table}

We also suspect that the coating of the cell was not optimal, but it worked adequately during the run. We will get a new cell that will be coated and tested following our previous approach \cite{aubin03a}.


A steady-state beam from ISAC delivers activity into the Y neutralizer in the down position for a fixed time between $10$ and $20$\,s (see figure~\ref{fig:MOT_sketch}). Pneumatic actuators operated with solenoid valves rotate the position of the neutralizer to the upper position for about $1.5$\,s. Once the neutralizer is up, a $\sim 8.9$\,A current flows through it and heats it for about $1$\,s to a temperature of less than 700 $^{\circ}$C.
The neutral francium is released and enters the glass cell for trapping. The new connection scheme for the neutralizer heating current (see figure~\ref{fig:alpha_det}a) worked well and reliably during $7000$~repetitions; before the run it was already tested without the final yttrium foil for about $25000$~cycles.

\begin{figure}
\begin{center}
  \includegraphics[width=0.85\linewidth]{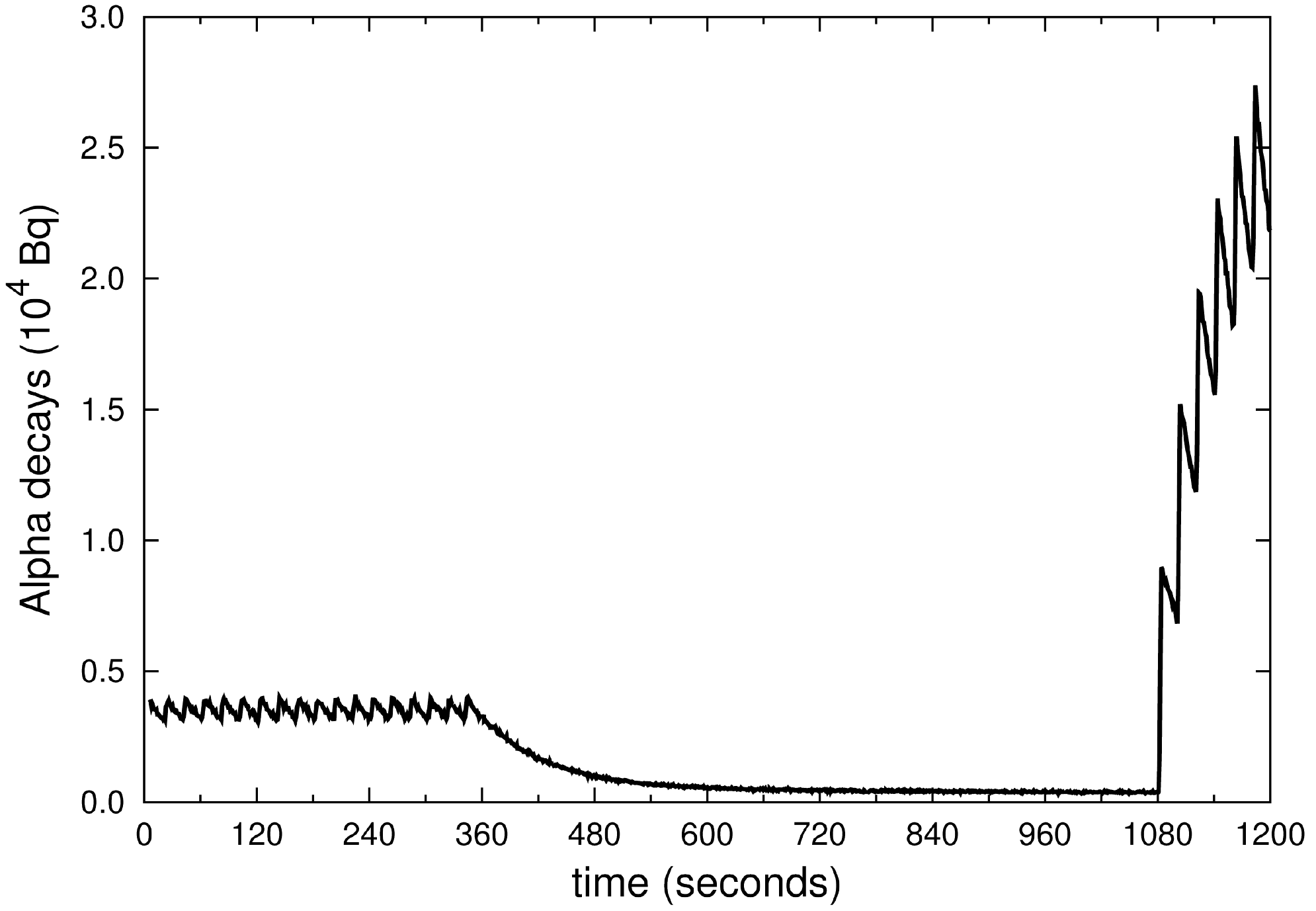}
\caption{Sample output of the $\alpha$ detector during the commissioning run. At the beginning of the time sequence $^{207}$Fr is being trapped. The `staggering' is caused by the yttrium foil blocking the beam most of the time. In the middle there is no radioactive beam. At the end of the time sequence (at $1080$\,seconds), $^{209}$Fr is delivered to the FTF. }
\label{fig:alpha_activity}
\end{center}
\end{figure}

The francium beam reaches the Faraday cup when the neutralizer is up. This allows us to monitor the $\alpha$ decays that give a direct measurement of the number of ions delivered. 
Figure \ref{fig:alpha_activity} shows such a signal which increases every $20$\,s. It also shows the change of isotope, from $^{207}$Fr to $^{209}$Fr, which was realized in about ten minutes. 
This same detector serves to measure the lifetime of the activity as it decays, which is dominated by the decay products of the corresponding Fr isotope.

\begin{figure}
\begin{center}
  \includegraphics[width=0.85\linewidth]{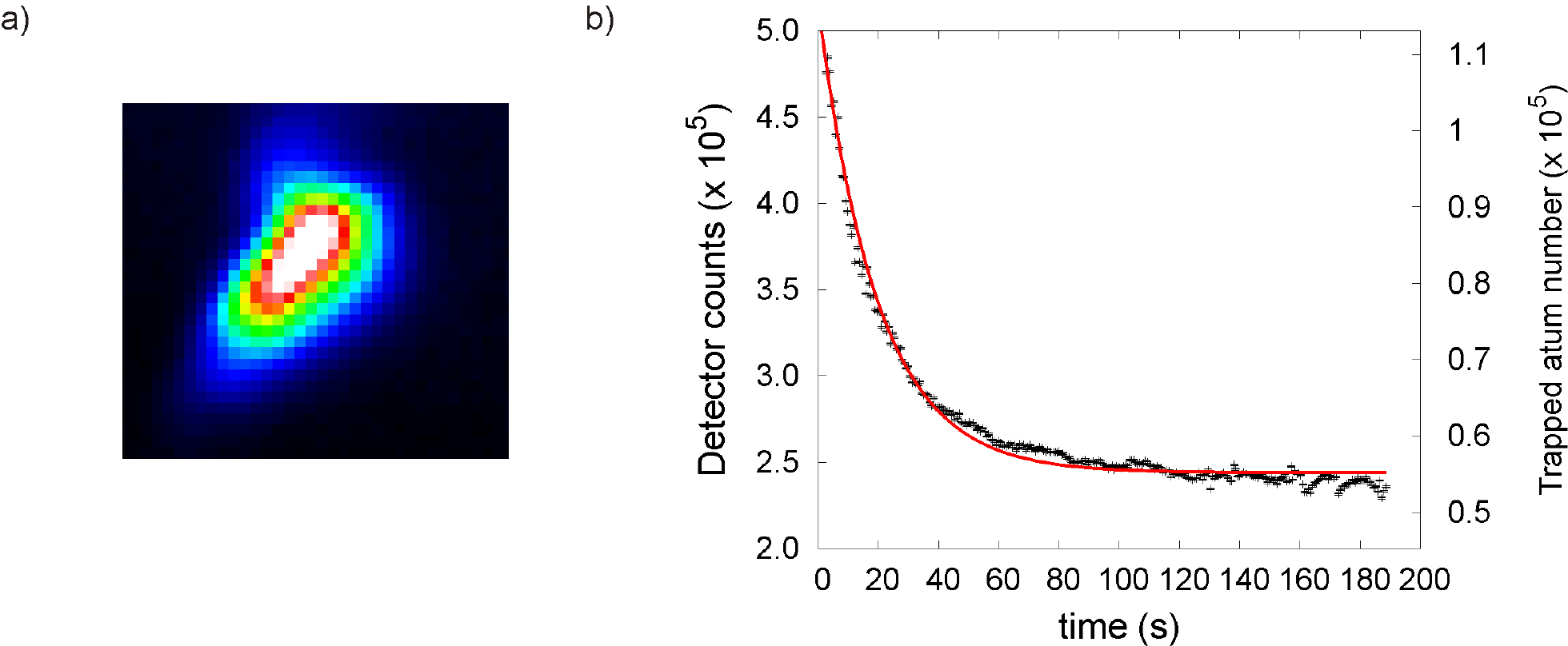}
\caption{Fr MOT performance. a) False color CCD image of the MOT fluorescence of a cloud of about $10^5$ $^{209}$Fr trapped at the FTF. The pixel size of the camera is $6.7 \times 6.7$\,$\mu$m$^2$; an area of $0.86 \times 0.86$\,mm$^2$ is shown.  b) Black data points: Integrated fluorescence from the trap as a function of time. Red line: An exponential fit results in a lifetime of about $20$\,s, even though there are deviations from this model in the data. The right y-axis is only approximate for these data as the calculated atom number has an uncertainty of $\sim 40$\,\%.}
\label{fig:Fr_trap}
\label{fig:trap_lifetime}
\end{center}
\end{figure}

Figure \ref{fig:Fr_trap}a) shows a false color image from the Roper Scientific camera that looks at the fluorescence at $718$\,nm. The size of the trap is about 500 $\mu$m in its largest dimension.
We do not apply a procedure to accumulate atoms in the capture MOT, as that would require sheltering the cold atoms from the hot burst of new Fr that arrive.
Tests off-line with Rb have shown that we can accumulate in a second chamber as long as the vacuum of the second chamber is sufficiently good \cite{perezPhD}. Accumulation can be done by having the MOT in the second chamber on -- with atoms from the previous cycle still trapped --, while a new batch of atoms is transferred from the capture MOT.

The software that processes the images requires as input the intensity of the trapping beams as well as the laser detuning to determine the number of trapped atoms. This serves to calculate the number of atoms by knowing how many photons are scattered per unit time. Independent tests with Rb using absorption imaging show good quantitative agreement with this way of measuring the number of atoms. While trapping Fr it is difficult to know the exact laser detuning to better than one half linewidth (4 MHz). This is in contrast with Rb (or any element with stable or long-lived isotopes) where the exact detuning is well-known since the laser frequency can be determined by comparison to a stable reference source.

\subsection{Different isotopes}
We report the successful trapping of $^{207}$Fr for the first time, based on the frequencies of the D1 and D2 lines measured at ISOLDE in the 1980's \cite{bauche86,coc85,coc87}.  $^{221}$Fr was used by the Boulder group \cite{lu97}, and we trapped it on-line to prepare for tests of an off-line Fr source \cite{tandecki13}.

\begin{table}[h]
\caption{Frequencies for trapping and repumping of the three working isotopes. The accuracy of $\pm 0.002$\,cm$^{-1}$ comes from the wavemeter and the fact that these are not the resonance frequencies. }
\begin{center}
\begin{tabular}{|cccc|}
\hline
Isotope    &  Trap (cm$^{-1}$)    &      Repumper - D2 (cm$^{-1}$)    & Repumper - D1 (cm$^{-1}$)  \\ \hline
207 &    13923.548   &    13924.945    &  \- \\
209 &   13923.470     &   13924.888    &  12238.380  \\
221 &    13922.952    &    13923.560   &   12237.054  \\ \hline
\end{tabular}
\end{center}
\label{isotope_parameters}
\end{table}

The frequencies for $^{209}$Fr (see Table~\ref{isotope_parameters}) agree with our previous work and with the work of the Legnaro group \cite{sanguinetti2009}. The frequencies for $^{221}$Fr agree with the trapping done by the Boulder group \cite{lu97}. The stability of the cavity while switching between isotopes enables us to measure isotope shifts with good accuracy using electro-optical modulator (EOM) sideband optical spectroscopy \cite{collister13}.

\subsection{Trapping results and efficiency}
\label{sec:efficiency}

We have successfully trapped up to $2.5 \times 10^5$ $^{209}$Fr atoms at the FTF. This gives an efficiency of about 0.05\%, which we calculate by taking the peak number of atoms in the trap and dividing it by the incoming francium beam integrated over the accumulation time. Fig~\ref{fig:trap_lifetime}b) shows the decay of a $^{209}$Fr trap that is not refilled; the lifetime of $20$\,s is the result of having good vacuum of the order of $10^{-10}$\,mbar. 
We found, however, that, despite several stages of differential pumping in our beamline, the ISAC low-energy beamlines were still influencing our trap lifetime depending on changes in environmental temperature (and thus water vapor pressure).

The reason for the difference of more than a factor of 20 from the best efficiency achieved at Stony Brook \cite{aubin03a} comes from three main factors: limited laser power, imperfect neutralizer heating, and the suspected low performance of the dryfilm coating.

The laser power available for trapping can be trivially increased. Our new laser system (M~Squared SolsTis) in combination with a high-power fiber will provide five times more intensity; i.e. from $\sim 200$\,mW going into the fiber to about $1$\,W. 
The efficiency from this upgrade is expected to increase by a factor of $15$; the laser light can be red-detuned further from resonance while maintaining the same photon scattering rate, and as a result the capture velocity, $v_c$, is increased \cite{lindquist92}.
From measurements of the release of francium after the ion beam was turned off, we can deduce that about $10$\,\% of francium atoms available in the yttrium foil were released on each heating cycle.
This could be improved by a factor of $5$ by heating the neutralizer more homogeneously to higher temperature to reach $50$\,\% \cite{aubin03a}.
Taking these two improvements into account, an efficiency of $3$--$4$\,\% becomes possible. 

\section{Outlook and conclusions}
\label{sec:conclusions_outlook}
The FTF is operating well. We have trapped up to $2.5 \times 10^5$ $^{209}$Fr atoms, which corresponds to an efficiency of $0.05$\,\%. The vacuum allows to have long trap lifetimes of about $20$\,s.

We are in the process of designing and constructing the science chamber and have identified the issues that need improving on the capture assembly. They include a new Y foil for neutralization with better electrical contact for more uniform heating, a new laser (M Squared SolsTis) in combination with a high-power fiber that delivers more trapping light.


We expect a transfer efficiency to the science chamber in Fr similar or better to the one reached with Rb ($50$\,\%) since Fr is heavier and has less transverse expansion.
The goal of $10^6$ atoms in the science chamber for the weak interaction measurements \cite{gomez07,sheng10} is within reach for our system.



\acknowledgments

We would like to thank the ISAC staff at TRIUMF for developing the Fr beam. NRC, TRIUMF, and NSERC from Canada, DOE and NSF from the USA, and CONACYT from Mexico support our work.


\end{document}